# Recommendations on the use and reporting of race, ethnicity, and ancestry in genetic research: experiences from the NHLBI Trans-Omics for Precision Medicine (TOPMed) program

Alyna T. Khan[1,2*], Stephanie M. Gogarten[1*], Caitlin P. McHugh[1], Adrienne M. Stilp[1], Tamar Sofer[3,4], Michael Bowers[1], Quenna Wong[1], L. Adrienne Cupples[5,6], Bertha Hidalgo[7], Andrew D. Johnson[8,9], Merry-Lynn McDonald[10,11], Stephen T. McGarvey[12,13], Matthew R.G. Taylor[14], Stephanie M. Fullerton[15], Matthew P. Conomos[1], Sarah C. Nelson[1,2]

1-Department of Biostatistics, University of Washington, Seattle, WA
2-Institute for Public Health Genetics, University of Washington, Seattle, WA
3-Department of Medicine, Harvard Medical School, Boston, MA
4-Division of Sleep and Circadian Disorders, Brigham and Women's Hospital, Boston, MA
5-Department of Biostatistics, Boston University School of Public Health, Boston, MA
6-Department of Epidemiology, Boston University School of Public Health, Boston, MA
7-Department of Epidemiology, University of Alabama at Birmingham, Birmingham, AL
8-Population Sciences Branch, Division of Intramural Research, National Heart, Lung and Blood Institute, Framingham, MA
9-The Framingham Heart Study, Framingham, MA
10-Department of Medicine, University of Alabama at Birmingham, Birmingham, AL
11-Department of Genetics, University of Alabama at Birmingham, Birmingham, AL
12-Department of Epidemiology and International Health Institute, Brown University School of Public Health, Providence, RI
13-Department of Anthropology, Brown University, Providence, RI
14-Department of Medicine, Adult Medical Genetics Program, University of Colorado Anschutz Medical Campus, Aurora, CO
15-Department of Bioethics & Humanities, University of Washington, Seattle, WA

*Co-first authors

# Abstract

The ways in which race, ethnicity, and ancestry are used and reported in human genomics research has wide-ranging implications for how research is translated into clinical care, incorporated into public understanding, and implemented in public policy. Genetics researchers play an essential role in proactively dismantling genetic conceptions of race and in recognizing the social and structural factors that drive health disparities. Here, we offer commentary and concrete recommendations on the use and reporting of race, ethnicity, and ancestry across the arc of genetic research, including terminology, data harmonization, analysis, and reporting. While informed by our experiences as researchers in the NHLBI Trans-Omics for Precision Medicine (TOPMed) program, the recommendations are broadly applicable to basic and translational genomic research in diverse populations. To fully realize the benefit of diversifying genetics research beyond primarily European ancestry populations, we as genetics researchers need to make structural changes to the research process and within the research community. Considerable collaborative effort and ongoing reflection will be required to root out elements of racism from the field and generate scientific knowledge that yields broad and equitable benefit.

# Introduction

Globally, we are reckoning with the structural racism embedded in nearly all aspects of our society, including health, health care, and biomedical research. Historically, the concept of race and racism has been preserved through scientific and medical advancements[1] and has persisted as a biologically-relevant concept in today's scientific thinking, illustrated by the recent attempts to use genetics to explain racial health outcome disparities of COVID-19[2]. However, race and ethnicity are social categories that divide people into groups based on social, political, and cultural norms, and while those categories can have biological effects on human health, they act via the embodiment of social and structural factors rather than innate genetic or biological difference[3,4]. Despite the recognition that both social determinants and genetic variation play a role in health outcomes, there is vigorous debate on the ultimate causes of health inequities, and ongoing disagreement about how to best integrate such information into genetic research.

One avenue to begin to address health inequities is to diversify genetics research. It is well-documented that the majority of genome-wide association studies (GWAS) have been conducted in populations of primarily European ancestry. This limited focus on a small subset of overall human genetic variation has led to a discovery bias that may exacerbate health and health care disparities and limit our understanding of human biology[5–10]. Sequencing non-European ancestry populations will enable us to observe more genetic variants, as it is well known, for example, that populations of African descent have the greatest amount of genetic variation[11]. Increasing diversity in genetics research will therefore improve our understanding of genetic variation and its impact on biological pathways, as well as make genetic information more broadly applicable to everyone.

While this focus on including more diverse populations in genetics research and addressing racial and ethnic health disparities is laudable and necessary, two serious harms are possible when such efforts are carried out without paying attention to the social and historical contexts underlying health disparities. The first harm is that focusing on genetic differences as an underlying cause of health disparities ignores the reality of racism as the most significant cause of these differences[12,13]. Some have argued that increasing diversity in genetics research will benefit people of non-European ancestry by revealing important genetic and biological variation *between* groups that have consequences on health[13–15]. Indeed, it has become common practice in GWAS of diverse populations to perform analyses stratified by race or ethnicity and to highlight that a particular disease or trait association is statistically significant in some strata of participants and not others. However, a potential unintended consequence of these group-based analyses is the implication that health disparities by race are due in large part to genetic differences, which obscures the influence of non-genetic (e.g., social and environmental) variables on health disparities. For example, in the case of COVID-19, many researchers published articles examining the possible genetic causes of the higher hospitalization and death rate in non-White people[2]. This avenue of research amounts to "blaming the victim," shifting the blame from racism to putative innate characteristics of non-White people and distracting from the most effective (societal) remedies[16]. Although there are oft-cited examples of differences in disease allele frequencies being strongly correlated with simplistic racial typologies (e.g., sickle-cell disease), cases in which genetics are the primary driver of race-based health differences are likely the exception rather than the rule[12,17]. Increased diversity in genetics research will improve our ability to search for new genetic variants that help us better understand biological mechanisms in everyone, but we must be careful not to assume that genetic differences associated with disease are a cause of the health disparity observed; rather, we must acknowledge the very real biological effects of racism on human health[4,18]. Calls for greater diversity and inclusion must also recognize and grapple with harms to racial and ethnic minority groups due to both past and ongoing practices in research and medicine[19].

The second harm of focusing on genetic differences between racial groups is that it contributes to racial essentialism[20], or the idea that race maps onto discrete genetic categories. Although racial essentialism has been definitively refuted by genetic research[21], its influence has not been eradicated. Indeed, empirical research of genomics professionals has found heterogeneous definitions and applications of race[22,23], including persistence of the notion that there is a genetic dimension to racial identity. In part due to this ambiguity, geneticists have attempted to move away from race as a biological concept by instead describing populations in terms of genetically-informed categories such as continental ancestry. However, in practice these categories are themselves closely aligned with racial and ethnic understandings[24–26]. Many genetic studies also either exert considerable effort to assign participants to discrete race, ethnicity, or ancestry categories, or exclude participants who do not fit into these categories[27]. Additionally, technological advances towards denser genotyping data have not made it any easier, or more appropriate, to define discrete populations based on genetic information, i.e. in the progression from ancestry informative markers (AIMs) to SNP arrays to whole genome sequencing (WGS). Given the vast numbers of people who do not self-identify with a single race and whose genomes are not neatly derived from a single ancestral continent[28,29], this is a failure of representation and a scientifically futile exercise.

Related to the problem of racial classification in genetic research, the NIH requires that studies collect data and report on the race of their participants according to the US Office of Management and Budget (OMB) census categories. Although the intent of this policy was to improve diversity in study participation and enable the elucidation of systemic racism by comparing outcomes between racial groups, in practice it can lead to race being used in causal inference as a biological rather than social category and other inappropriate contexts[30]. Scientists continue to demand a shift in practice[31], urging the scientific community to move away from using race as a measure of biological difference, and instead, use terms like "ancestry" or "population" to describe groups, which should be clearly and explicitly defined, and should avoid simply serving as a rebranding of race. Race is also commonly used to capture the effects of racism on health, and scientists are encouraging the use of a measure of "racism" instead of race to address how structural racism impacts health[32,33]. Recently, there have also been calls for reforming the terminology used in genetics scholarship in an attempt to disambiguate race and genetic ancestry[34]. Given the deep correlation (but not causation) between these two concepts, achieving this goal will require dedicated effort from all genetics researchers.

# Background

As study investigators, analysts, and support staff working with the NHLBI Trans-Omics for Precision Medicine (TOPMed) program, we are motivated to conduct scientifically robust and ethically responsible genetic research that benefits everyone and avoids the potential harms noted above. To that end, we have developed recommendations for the use and reporting of race, ethnicity, and ancestry in TOPMed, which are broadly applicable to genetic research in diverse populations. Below, we describe the formation of these recommendations in more detail and provide context for their development and promulgation across TOPMed. These recommendations are presented in the following section and are summarized in [Box I](Box I).

## TOPMed as a motivating use case

TOPMed is a large consortium of ongoing 'omic' (i.e., genomic, transcriptomic, proteomic, metabolomic, and methylomic) studies that encompass people of many different races, ethnicities, geographic locations, and ancestries[35]. TOPMed includes studies based within and outside of the US, in addition to unique founder populations such as Samoan and Amish. This diversity of study populations is a major strength of TOPMed; it enables the expansion of knowledge of genetic variation and an improved understanding of disease[36]. For example, of 400 million TOPMed variants reported by Taliun et al.[35], 78.7% were not previously deposited in dbSNP. However, the diversity in TOPMed also presents numerous research challenges, including harmonizing demographic information, conducting association analyses, and reporting research findings. Addressing these challenges is necessary to fully realize the ability of the TOPMed program to contribute robustly and equitably to precision medicine.

## Early development at the TOPMed DCC

The TOPMed Data Coordinating Center (DCC) is housed in the Genetic Analysis Center (GAC) in the Department of Biostatistics at the University of Washington—a group that has performed scientific, analytical, and/or administrative coordination for a range of human genomics consortia and programs over the past 15 years including the NHGRI Gene Environment Association Studies consortium (GENEVA, 2007-2011), NHGRI Genomics and Randomized Trials Network (GARNET, 2009-2012), and the NHLBI Hispanic Community Health Study/Study of Latinos (HCHS/SOL, 2013-2016). Through these efforts, we have established standards for genotypic data quality assurance that account for population structure[37], grappled with how to analyze and report genetic diversity, e.g. among Hispanic/Latino groups[38], and developed statistical methods and software for analyzing diverse datasets[39–42]. In 2018, the GACinitiated monthly internal discussions on the use of race, ethnicity, and ancestry in genetics research—engaging with academic literature, public media, and our own experiences working in TOPMed and prior genetics consortia. This laid the groundwork for developing recommendations in TOPMed. We discussed a variety of articles across disciplines[43–46] and invited guest speakers on topics such as statistical rationale for stratified analyses and the co-opting of population genetic research by white supremacists on social media[47]. Discussions were informed by a range of training and experience at the DCC, including biostatistics; statistical genetics; science communication; public health genetics; and ethical, legal, and social implications (ELSI). From these discussions, we recognized the opportunity as a DCC to establish some recommendations for TOPMed researchers that address the challenges of working with diverse data and incorporate antiracist principles into the research process.

## Establishing recommendations for TOPMed

The motivations for compiling a set of recommendations for TOPMed investigators were two-fold: (1) to encourage researchers to make well-founded and responsible analytical and methodological decisions when using race, ethnicity, and ancestry variables and (2) to communicate concepts of race, ethnicity, and ancestry in an informed, transparent, and respectful manner. After initial drafting at the TOPMed DCC, these recommendations were discussed in relevant TOPMed Committees (ELSI and Analysis), approved by the TOPMed Executive Committee, and presented at Consortium-wide meetings. We solicited examples from study investigators of study-specific considerations and preferences, e.g. for population labels, and sought to incorporate diverse expertise and experiences to make the recommendations a practical, robust, and compelling resource to a wide audience of genetics researchers. Ultimately, these recommendations aim to help investigators navigate some of the challenges in using socially and genetically defined groups in scientific discussions by presenting an overview of commonly used terminology, highlighting considerations for data harmonization and analysis, and discussing matters in reporting of findings. They engage with and build on similar recommendations issued by leaders in biomedical research and publishing[34,45,48,49]. While developed in the context of the TOPMed program, we contend that these recommendations are relevant for genetic and biomedical researchers working in other contexts, especially those involving diverse populations and/or the genetic study of conditions that suggest health disparities.

# Guidelines

## Terminology

When presenting information on the race, ethnicity, or ancestry of participants in a study, it is essential to be clear about whether the labels used refer to reported or genetically inferred information. People may use these terms in different ways, and even among dictionaries there is no clarity on their precise meaning. "Race" and "ethnicity" generally refer to social, not biological, categories, and they are often used interchangeably, or as the hybrid term "race/ethnicity." In contrast, "ancestry" is generally used in genetic research to refer to one's biological ancestors from whom their DNA was inherited, or to imply something about a person's genetic origins; for example, whether the majority of their ancestors were from Africa, the Americas, Europe, or Asia (sometimes referred to as "continental ancestry")[45,50]. Ancestry can also be described on a finer scale, such as having ancestors from specific countries or geographic regions, and this is how "ancestry" is often used colloquially in non-scientific settings. In these recommendations, we use the terms "race" and "ethnicity" to refer to non-biological social categories, and we use the term "genetic ancestry" to describe genetic origins. Because reported race or ethnicity and genetic ancestry may all be used analytically and appear in scientific discussions and communications, care must be taken to describe exactly what is being presented and why.

Recommendations for investigators:

1. **Explicitly distinguish between variables that derive from non-genetic, reported information versus genetically inferred information.**
2. **Avoid using terms that are historically linked to hierarchical, racial typologies.** For example, the term "Caucasian" should not be used[3,51]; instead, use "White" when referring to race and "European ancestry" when referring to genetic ancestry.
3. **Follow standards from publishers**, including the APA's guidelines on bias-free language regarding racial and ethnic identity and the AMA Manual of Style (see Flanagin et al. 2021).

## Harmonization of Race and Ethnicity Across Studies

Race and/or ethnicity are often collected by a study and included with other phenotypes describing study participants (such as sex, age, and height). A common method of collecting this information is for study participants to fill out a form indicating their race and/or ethnicity (typically choosing one or more from among a set of options provided), which leads to "self-reported" values. Other collection methods are also possible, including designation by a third party (health care provider or study data collector) who typically infers the participant's ascriptive race, or through study documents that describe the recruitment population but do not ask whether the self-reported race and/or ethnicity of specific individuals differs from the target population. However collected, the race and/or ethnicity of a participant is almost always a function of the specific options provided in study instruments, which will often vary by location or the research interests of investigators. Additionally, if collected longitudinally, a participant's self-identification may change over time, e.g. if their perceptions of race, their own identity, or their family history change.

The diversity in data collection methods presents a challenge for investigators attempting to combine data from multiple studies. Unlike quantitative phenotypes measured with different units that can be transformed to the same scale during data harmonization, there is often no straightforward method to convert one set of race or ethnicity categories into another. This is particularly the case when study cohorts include individuals sampled from distinct national contexts where socio-cultural understandings of racial and/or ethnic identity differ, when working with studies with very different recruitment periods, or when different studies provide different options for race and ethnicity categories (such as offering the descriptor "Asian" on a form versus offering more specific identifiers, like "East Asian" or "South Asian"). It is important to keep in mind the complexities and nuances of social identity when attempting to harmonize race and ethnicity variables across studies.

Recommendations for investigators:

1. **Clearly describe the source data from each study when using harmonized race and ethnicity variables.** Include details such as whether source information is self-reported or ascribed, and whether multiple categories are collapsed into one. Be aware that cross-study harmonized variables often represent a simplification of more complex sources of information that may not translate well between different studies and jurisdictions (e.g. different countries, recruitment periods, or specificity of provided options).
2. **Avoid assuming that non-genetic, reported variables are by "self-report."** Study- or cohort-specific documentation may help determine whether variables (e.g. race or ethnicity) were self-reported versus recorded by study personnel without soliciting self-report from the participant.
3. **Avoid applying US race categories to participants of studies based outside of the US.** Concepts of racial and/or ethnic identity differ across countries, and approaches to capturing this information vary across geographic location and over time[52]. Some countries do not collect race information at all; for example, Australia abandoned the use of racial classification in 1974, and instead collects information on ethnicity[52]. Brazil, however, collects race information, but racial categories in Brazil differ from those of the United States[53,54]. Someone who might identify and/or be identified as Black in the US will not necessarily consider themselves and/or be considered Black in Brazil.
4. **Consult study documentation or ask study representatives how their study participants prefer to be described.** Because notions of racial and ethnic identities vary in different contexts, it is important to describe participants according to their preferences and in ways that reflect their social and cultural contexts.

## Genetic Ancestry

Genetic ancestry can be inferred indirectly from measured genotype data by examining genetic similarity, either among participants with reported ancestry information, or to reference samples of known ancestry[50]. Principal component analysis (PCA) is commonly used to describe the variation in genotype data as a continuous, multidimensional distribution, with participants whose ancestors came from the same geographical area often clustering together in PC space[55]. Admixture analysis is also commonly performed to estimate the proportion of each participant's genome descended from pre-specified reference populations of known ancestry. The continuous nature of these genetic ancestry PCs and ancestry proportion estimates illustrate the heterogeneity in genetic ancestry among individuals who may identify as the same race or ethnicity, particularly in admixed populations, defined as groups of people whose genetic ancestry derives from multiple previously isolated populations. For example, those who identify as Hispanic/Latino have a wide variety of genetic ancestries, with different proportions of ancestry admixture from Africa, the Americas, Asia, and Europe[38,56,57]. Simply using race and/or ethnicity as a proxy for genetic ancestry is problematic in that it falsely equates the two correlated, albeit distinct, concepts.

Recommendations for investigators:

1. **Avoid using reported race or ethnicity as a proxy for genetic ancestry, or using genetic ancestry to represent race or ethnicity.** Race and ethnicity can be correlated with genetic ancestry, but they are not the same. Individuals who identify as the same race or ethnicity can have a wide variety of genetic ancestries, and individuals with similar genetic ancestry may identify as different races or ethnicities.
2. **Avoid reinforcing the idea that race and ethnicity are genetic concepts when presenting genetically derived data.** When presenting figures or summary statistics, be clear about how labels were defined, use terms that represent the source of the information, and justify their use in the given context. For example, if labeling participants in PC plots by their race and/or ethnicity, say why this was done, and use the original racial or ethnic designations rather than re-labeling with (proxy) ancestry terms. As another example, do not assume that allele frequencies from a reference population apply to a particular racial or ethnic group, or vice versa; e.g., the allele frequencies in an African-American population are not the same as those in the HapMap Yoruba in Ibadan, Nigeria (YRI) population.

## Analysis

When considering how to use race, ethnicity, and/or genetic ancestry information in an analysis, analysts should first assess the goals of the study and the intended purpose of inclusion of those variables in models, as they would with inclusion of any variable. In GWAS, the goal is to identify genetic variants that are associated with a particular trait or disease, and race, ethnicity, and genetic ancestry can all be relevant to consider in such an analysis. Genetic ancestry may be a confounding factor in the analysis if it is associated with the trait or disease of interest, as allele frequencies and patterns of linkage disequilibrium at many variants differ between populations[58]. We discuss below some considerations when adjusting for confounding due to genetic ancestry. Race and ethnicity are often tied to social factors influencing health; examples include racially-based housing discrimination which influences environment[59–61], and increased stress levels in individuals experiencing racism[62,63]. In addition, reported race and ethnicity may explain variation in the trait or disease of interest that is dependent on aspects of social identity (e.g. systemic or individual racial discrimination), rather than genetic ancestry. For example, African Americans with a high proportion of European ancestry may suffer the same lack of access to adequate health care as African Americans with little to no European ancestry[64]. As another example, diet is correlated with many health outcomes and is often culturally or socioeconomically driven[65]. In such instances, when the underlying variable(s) of interest (e.g. measures of health care or diet) are unavailable, including race or ethnicity as a covariate may improve statistical power to detect association.

One of the most popular approaches to address confounding due to genetic ancestry in association tests is to conduct a meta-analysis, where different racial, ethnic, or ancestry groups are stratified and analyzed separately, and summary statistics from each group are subsequently combined for inference on the entire study. Meta-analysis can be effective at controlling for confounding; however, we encourage investigators who take this approach to focus on the final meta-analysis results and exercise caution when interpreting the group-specific results. A commonly referenced motivation for interpreting the group-specific results is to determine whether participants of a particular ancestry are "driving" the observed association signal. While a statistically significant association may be observed in one group and not another, in our experience, this often appears to be a result of differences in statistical power to detect an association (e.g. due to sample size or allele frequency differences), rather than fundamental differences in the underlying biological impact of the same variant in different groups of people.

Another commonly used approach to adjust for confounding due to genetic ancestry is to perform a pooled-analysis (i.e., an analysis including all study samples) and include PCs calculated from sample genotype data as covariates. These PCs typically capture genetic variation among study participants due to ancestry, including admixture, on a continuous scale and effectively control for confounding. An alternative but conceptually similar approach to using PCs is to include as covariates ancestry proportions for each subject, estimated using reference samples of known ancestry and software such as ADMIXTURE[66]. A distinct advantage of this pooled-analysis approach over a stratified-analysis approach is that it does not require arbitrary clustering decisions or cross-study harmonization of demographic variables and also allows inclusion of all participants in the analysis, including those with either missing or underrepresented race or ethnicity[27].

When analyzing TOPMed data, we typically do not find additional signals from considering group-specific analyses that are not also identified by pooled-analysis including the same individuals. On the other hand, population-specific results of previously understudied populations may provide actionable findings for that population. Therefore, it is critical to engage with study participants or representatives on whether it is appropriate to pursue population-specific analysis and how best to represent them in the study. Ultimately, it is important to recognize the various technical and contextual factors that influence analytical decisions and to be transparent about which approach was taken and why.

Recommendations for investigators:

1. **Articulate and justify why variables were used in a given analysis.** In particular, explain the reasoning behind analytical decisions to use non-genetic and/or genetically inferred variables, e.g. in methods sections. Analytical decisions are nuanced and often reflect a weighing of various pros and cons to different approaches.
2. **Keep in mind that, if using race or ethnicity as a covariate, these variables may explain trait variation due to social factors, not genetics.** In most cases, association between reported race or ethnicity and allele frequencies is due to a correlation between race or ethnicity and genetic ancestry, so the inclusion of PCs or ancestry proportions as covariates is usually sufficient to account for confounding. Race or ethnicity may correlate with non-genetic, social factors, but the effects of such factors can be better accounted for when used directly, if the data are available.
3. **Focus attention on pooled- or meta-analysis results of all participants.** Whether a pooled-analysis or a meta-analysis is used may depend on technical limitations—e.g. data sharing constraints—and the ability to jointly harmonize and analyze the genetic and phenotypic variables across studies. Describe which approach was taken, why, and what the limitations may be. If considering group-specific results, keep in mind that individuals within racial or ethnic groups do not have homogeneous genetic ancestry, and avoid reinforcing an equivalence between race or ethnicity and genetic ancestry.
4. **Consider potential benefits versus potential harms when thinking about whether and how to conduct a population-specific analysis.** Consult with study representatives or documentation to understand if their study participants would find it acceptable, or even preferred, to acknowledge their unique population history and evolution. For some understudied populations, population-specific results may provide actionable findings for that population[66,67]. However, in some instances, participants may not wish to associate membership in their population with a specific trait that could be considered stigmatizing[68].

## Reporting

Reporting on race, ethnicity, and ancestry variables and constructs is typically necessary to describe one's approach and methods as well as to provide interpretation of results. Past studies have found inadequate descriptions of race, ethnicity, and ancestry variables in the scientific literature[49,69], which can lead to both scientific and social harms (see Introduction). Below, we offer some recommendations on the reporting of race, ethnicity, and ancestry variables. Notably, these recommendations are meant to augment rather than supplant existing and emerging reporting recommendations from journals and funders[e.g.34,48].

Recommendations for authors or presenters:

1. **Acknowledge the broader social context of health and healthcare disparities when invoking these disparities as a justification for genomic research.** Health disparities are differences in health "linked with economic, social, or environmental disadvantage"[70]. While health disparities often disproportionately affect minority racial and ethnic groups, the underlying reasons are typically due to social and structural determinants of health rather than genetic factors[12,71,72]. Genetic research may be part of the solution to address health disparities, but should be integrated into "social models of disease and interdisciplinary research methods"[16].
2. **Consult with study investigators or study-provided documentation about any preferences or study-specific reporting guidelines.** Given the number and complexity of studies with diverse data, and the potential for conflicting study-specific recommendations in cross-study analyses, we strongly encourage authors to discuss these issues with study representatives. Where direct access to original study investigators or participant representatives is infeasible, expend effort identifying reporting standards or precedents in the study.
3. **Avoid generalizing from a single population to represent another, broader population.** Keep in mind the limitations of population identifiers and generalizability to larger population groups[73]. For example, if a study includes Samoans but no other Pacific Islander populations, do not generalize the Samoan people to represent all Pacific Islanders.

## Conclusion

Conducting genetic research in the context of large-scale, diverse consortia presents both challenges and opportunities. For example, the TOPMed program comprises over 80 contributing studies with diversity in terms of populations, geographic locations, genetic ancestries, and areas of phenotypic focus. To make use of the strengths and benefits afforded by diversifying genetics research, we as genetics researchers need to make structural changes to the research process and within the research community. We should critically evaluate each step of the research process, from hypothesis generation and study design, to data collection, harmonization, analysis, and reporting, to ensure that unintended or hidden misuses of race are rooted out. For example, when we set out to identify genetic associations with disease and explore whether differences in association between racial groups exist, it can be easy to conclude that it must be these genetic differences that are driving the outcomes we observe rather than social or structural determinants of health such as racism. One way to curtail this conclusion is to begin by stating an explicit hypothesis that calls attention to the social and environmental factors contributing to the outcomes observed[74]. Additionally, measuring key social and structural factors and integrating those into genetic analyses can help elucidate environmental contributions and gene-by-environment interactions. It is important that, when studying differential health outcomes or group differences, we counteract rather than reinforce racialized thinking. Here, we have offered recommendations based on our experience in the TOPMed consortium intended to contribute to this larger structural change and reimagining of genomics research moving forward. Below we situate these recommendations in emerging areas of genetics research, broader conversations on reporting standards and methodological advances, and ongoing calls to diversify research participants and the genetic workforce.

    We recognize our recommendations as part of a broader conversation in the scientific community about refining reporting guidelines and advancing statistical and other research methodologies needed to strive for an anti-racist science[31]. For example, establishing new standards for terminology and incorporating updated publication requirements that demand clear and rigorous definitions of race, ethnicity, and ancestry variables are crucial in extinguishing racialized thinking from genetics research and literature[33,34], as it encourages investigators to be more critical when applying these concepts in the design, development, and conduct of their research. In addition to changes in language and reporting, methodological advancements and a reevaluation of existing methodologies are necessary to continue improving how researchers use and define these concepts. For example, systematic investigation of taking a stratified versus pooled approach to association testing will provide empirical evidence for if, and when, stratifying participants is necessary. This work is sorely needed because, if used indiscriminately, stratification by race may reify race as genetic and obscure the non-genetic, "fundamental causes" of health inequities[75]. Approaches that reinforce biological concepts of race may ultimately harm—or minimally fail to help—the underserved communities that genetics research seeks to include. Methodological advancements are also necessary to accommodate analyses of diverse populations, where standard methodologies developed under an assumption of homogeneous populations may not be appropriate[76].

    We should also critically examine the widespread use of "continental ancestry" in genetic research. The selection of reference populations with ancestry from specific geographic areas is somewhat arbitrary, yet these samples are widely used to represent entire continents[77]. For example, despite early guidance against such oversimplification[73], the HapMap Yoruba in Ibadan, Nigeria (YRI) are often used to represent all of Africa; however, this population represents a small amount of diversity present across African genomes. Further, the usual classification of people as having "European", "Asian", "American", or "African" ancestry makes reference to a specific time period, i.e. after the global geographic dispersal of *Homo sapiens* from Africa and prior to the European colonization, especially of the Americas, that accompanied the so-called Age of Discovery. We could just as easily define continental ancestry based on a different time period, such as current human geography[77], and it would be no more right or wrong, but would lead to a very different understanding of, for example, "American"

ancestry. While categorizing ancestry components by continent can be a useful model of the data, we must keep in mind that it is only a model, and one that obscures genetic heterogeneity within continents and the complex, dynamic political, social, and migratory histories of those regions[78]. As scientists, we are trained to evaluate new data to see if it matches what we expect to be true. This training can work against us when it intersects with our social biases, since we view results that reflect those biases as more likely to be "true" than other results. This can lead to a belief in the correspondence of continental ancestry with historical "races" rather than recognizing the practice of clustering genomes in more or fewer population groups as a modeling choice[4]. Allele frequencies and patterns of linkage disequilibrium are known to differ across populations, but these differences are a result of processes including mutation, genetic drift, selective pressure, and population bottlenecks and expansion, reflecting rich population history and migration[77]. Variations in allele frequencies do not indicate natural, static genetic differences between a fixed number of population groups.

Finally, efforts to improve diversity in genetics research must be heeded. Diversifying genetics research not only means increasing diversity of study populations, but also improving and supporting diverse representation within the genetics research community itself. For example, there has been much discussion about how diverse representation among genetic scientists might help overcome barriers of mistrust between minority groups who have been historically exploited in biomedical research studies and the scientific community[79,80]. A commitment to making genetics research equitable and applicable to all means we must also be committed to recruiting, supporting, and amplifying the voices of underrepresented scientists in academia and the genetics community more broadly.

Averting and correcting misuses of race and ancestry in genetics research now is critical before it potentially gets "baked into" emerging applications. For example, the development of polygenic risk scores (PRS) is an emerging area in genetics in which concerns about the availability and use of diverse datasets are especially pertinent. PRS provide estimates of an individual's risk for a discrete, clinically relevant outcome and are a subset of polygenic scores (PGS), which quantify aggregate genetic predisposition to a trait[81] (for simplicity, we refer to "PRS" moving forward). PRS are typically based on summary statistics derived from GWAS data. We are facing several challenges in the implementation and interpretation of PRS, such as poorer predictive performance in non-European and admixed individuals due to the over-representation of European populations in GWAS, which could, if left unaddressed, exacerbate health and healthcare disparities[82]. As more attention is given to the development and improvement of PRS in diverse populations (e.g., [NOT-HG-20-010](NOT-HG-20-010)), we must critically evaluate the roles that race, ethnicity, and ancestry play in this effort. Diversifying study populations in GWAS is of prime importance for improving the applicability of PRS to non-European and admixed populations and avoiding further disparity. Further, we believe that the recommendations that we have presented here remain applicable to PRS development and application, as an extension of GWAS research. Thinking critically about the assumptions we make behind our analytical choices and methodological approaches will be paramount to preventing racial essentialism from clouding our understanding of PRS and the role that genetics plays in complex disease.

Ultimately, awareness, transparency, and sensitivity among researchers is needed to encourage thoughtful data stewardship, foster collaboration, and work towards expanding the diversity and representation needed to further translational genomic research[6]. As genetic scientists, we have a responsibility to ensure that we promote meaningful genomic knowledge and scientific advancements that benefit everyone. We recognize that addressing race, ethnicity, and ancestry in genetics research is a nuanced practice with changing perspectives. There is much to learn on how best to appropriately consider social factors in genetics research and translation and ensure that we dismantle any remnants of racialized thinking from this work. In order to tackle these issues successfully, we must be open to new and evolving ideas and approach this work with ongoing reflection and humility.

# Acknowledgements

The Trans-Omics in Precision Medicine (TOPMed) program is supported by the National Heart, Lung and Blood Institute (NHLBI). The Data Coordinating Center at the University of Washington has been funded under R01-HL120393, U01-HL120393, and contract HHSN268201800001I. STM acknowledges grant support from R01-HL093093 and R01-HL133040. ADJ acknowledges NHLBI Intramural Funding. We are grateful to the GAC Race and Genetics Discussion group and members of the TOPMed ELSI (Ethical, Legal, and Social Issues) Committee, who provided valuable feedback and input on earlier versions of this work. The views expressed in this manuscript are those of the authors and do not necessarily represent the views of the National Heart, Lung, and Blood Institute; the National Institutes of Health; or the U.S. Department of Health and Human Services. We gratefully acknowledge the studies and participants who provided biological samples and data for TOPMed.

# Authorship Contribution

Conceptualization: A.T.K, S.M.G, C.P.M, M.P.C, and S.C.N.; Writing - Original Draft: A.T.K, S.M.G, C.P.M, M.P.C, and S.C.N; Writing - Review and Editing: A.T.K, S.M.G, C.P.M, A.S, T.S, M.B, Q.W, L.A.C, B.H, A.D.J, M.M, S.T.M, M.R.G.T, S.M.F, M.P.C, and S.C.N.

**Box I. Summary of recommendations on the use and reporting of race, ethnicity, and ancestry in genomics research.**

1. **Terminology**
   1.1. Explicitly distinguish between variables that derive from non-genetic, reported information versus genetically inferred information.
   1.2. Avoid using terms that are historically linked to hierarchical, racial typologies.
   1.3. Follow standards from publishers, including the APA's guidelines on bias-free language regarding racial and ethnic identity and the AMA Manual of Style (see Flanagin et al. 2021).
2. **Harmonization of race and ethnicity across studies**
   2.1. Clearly describe the source data from each study when using harmonized race and ethnicity variables.
   2.2. Avoid assuming that non-genetic, reported variables are by "self-report."
   2.3. Avoid applying US race categories to participants of studies based outside of the US.
   2.4. Consult study documentation or ask study representatives how their study participants prefer to be described.
3. **Genetic Ancestry**
   3.1. Avoid using reported race or ethnicity as a proxy for genetic ancestry, or using genetic ancestry to represent race or ethnicity.
   3.2. Avoid reinforcing the idea that race and ethnicity are genetic concepts when presenting genetically derived data.
4. **Analysis**
   4.1. Articulate and justify why variables were used in a given analysis.
   4.2. Keep in mind that, if using race or ethnicity as a covariate, these variables may explain trait variation due to social factors, not genetics.
   4.3. Focus attention on pooled- or meta-analysis results of all participants.
   4.4. Consider potential benefits versus potential harms when thinking about whether and how to conduct a population-specific analysis.
5. **Reporting**
   5.1. Acknowledge the broader social context of health and healthcare disparities when invoking these disparities as a justification for genomic research.
   5.2. Consult with study investigators or study-provided documentation about any preferences or study-specific reporting guidelines.
   5.3. Avoid generalizing from a single population to represent another, broader population.